\documentclass{epl}
\title{Effect of temperature and bias voltage on the conductance
distribution of disordered 1d quantum wires}
\shorttitle{Effect of temperature and bias voltage}
\author{Victor A. Gopar\inst{1,2}\thanks{Present address: Instituto de 
Biocomputaci\'on y F\'{\i}sica de los Sistemas Complejos, Corona de 
Arag\'on 42, 50009 
Zaragoza, Spain} \and Peter W\"olfle \inst{1}}

\institute{                    
  \inst{1} Institut f\"{u}r Theorie der Kondensierten Materie,
             Universit\"{a}t Karlsruhe, Wolfgang-Gaedestr. 1, 
             76128 Karlsruhe, Germany\\
  \inst{2} Max-Planck-Institut f\"ur Physik komplexer Systeme, N\"othnitzer
  Strasse 38, 01187 Dresden, Germany
}

\begin{document}
\maketitle
\begin{abstract}
The statistical properties of the conductance of one dimensional
disordered systems are studied at finite bias
voltage $V$ and temperature $T$, in an independent-electron picture.  We 
calculate the complete distribution of the
conductance $P(G)$ in different regimes of $V,T$ within a statistical
model of resonant tunneling transmission. 
 We find that $P(G)$ changes
from the well-known log-normal distribution at $T=0$ in the linear
response regime to a Gaussian distribution at large $V,T$.  The
dependence on $T$ and $V$ of average quantities such as 
$\langle G \rangle,\; \langle \ln G \rangle $ is analyzed
as well.  Our analytical results are confirmed by numerical
simulations. 
We also discuss the limits of validity of the model and conclude that
the effects of finite $T, V$  presented here should be observable.
\end{abstract}

Quantum electronic transport in one-dimensional disordered systems has
been widely studied ever since the concept of Anderson localization
has been introduced in the sixties \cite{mesoscopic}.  In particular 
it is known that
within the model of noninteracting electrons all states are localized
in 1d, with average extension $\xi$, the localization length.  For
systems of finite length $L$, the typical conductance decreases
exponentially with increasing length.  However, there are large sample
to sample fluctuations and a statistical analysis of the conductance $G$
is required.  For a 1d wire of length $L \gg \xi$, the full
distribution of linear conductance at $T=0$ has been
determined \cite{mesoscopic,pier,rejaei,beenakker, muttalib}: 
it is log-normal with a large variance of the order of the
mean.

Finite temperature $T$ and a finite bias voltage $V$ have a dramatic
effect on the conductance distribution, as we show in this letter.  
The conductance is given by an energy integral of the
transmission coefficient $g(E)$. A window of energies within which electron 
transport is possible is determined by the value of $T$ and $V$. Since $g(E)$ 
is a strongly fluctuating function of $E$, the result depends sensitively on 
the width of the energy window.

In the linear conductance regime (zero bias voltage), Azbel and Vincenzo
\cite{azbel1} 
calculated the $T$ dependence of the
averages of the conductance and its logarithm, at low temperatures. They
employed a model of $g(E)$ as 
a sum of resonances with vanishing width ($\delta$-functions) caused
by resonance tunneling through localized states situated at random
positions, and decaying exponentially over the distance $\xi$. They 
did not consider the full distribution of conductances.
Moreover, they did not take into account
the fluctuations of the localization length. Their model therefore
does not recover the log-normal distribution of $G(T)$ at $T=0$. More  
recently, it has been shown numerically that the average
resistance decreases strongly with increasing temperature \cite{mosko}.

The statistical properties of the nonlinear conductance
$G(T,V) = J(T,V)/V$, in particular its distribution $P(G)$ have not
been considered before (here $J(T,V)$ is the charge current), to our
knowledge. 

One may ask whether interaction effects will preempt the behavior found
here within the independent particle model.  The phase coherence
required for resonant tunneling is limited to a finite length
$L_{\phi}$ by phase relaxation at finite temperature/voltage.  We will
argue below that for sufficiently short wires $L \ll L_\phi$, even at
the values of $T,V$ of interest and phase coherence is preserved.
Furthermore, at finite $T,V$
transport will occur by way of inelastic processes, {\it e.g.} thermally
activated hopping \cite{mott,stone,serota}.  As shown below, these 
processes will dominate the 
transport in a regime of $T,V$ beyond the one we will consider here. 
Our estimates show that the model we 
consider should be applicable to realistic systems.

In this letter we study the effect of bias voltage and temperature on
the statistical properties of transport in a one-dimensional
disordered system, within the model of independent electrons.  We
consider transmission through the wire by way of resonant tunneling,
and employ a statistical model of individual resonances with Breit-Wigner 
line shapes \cite{foxman}.  We calculate
the full distribution of the conductance in various regimes of $T,V$,
as well as average quantities such as $\langle G \rangle $ and 
$\langle \ln G \rangle$.

We show that the distribution of the conductance develops from a log-normal
distribution in the limit of zero voltage and temperature to a 
Gaussian distribution 
in the high $V$ and $T$ regime. We also find that the conductance 
average $\langle G \rangle$ is independent of $V,T$, while 
$\langle - \ln G \rangle$ decreases with $V,T$ from its zero temperature and
voltage value
up to $- \ln \langle G \rangle$, at high values of $V,T$.

Our analytical 
results are supported  by numerical simulations of a disordered 
one-dimensional 
multibarrier system. We calculate numerically the total scattering matrix  
of the system by combining the scattering matrices associated to each
barrier. We consider all the barriers to have the same fixed height, 
but random widths and separations \cite{units}.

\section{The model} The nonlinear conductance $G(T,V)$ in units of 
$G_0 = 2e^2/h$ is given by \cite{bagwell,datta}
\begin{equation}
\label{gtv}
G(T,V) = -\frac{1}{eV}\int_{-eV/2}^{eV/2}dE{'} \int_{-\infty}^{\infty} dE 
g(E)\frac{\partial f(E-E')}{\partial E} ,
\end{equation}
where $f(E) = \left(\exp [(E-\mu)/kT] + 1\right)^{-1}$ is the Fermi 
function; $k$, the
Boltzmann  constant; and $\mu$, the
chemical potential.  
$g(E)$ is the transmission probability, equal to the zero
temperature linear conductance for $E=E_F=\mu(T=0)$.  In a wire
of length $L \gg \xi$, where $\xi$ is the average localization
length, transmission through the wire takes place via localized states
with random energies $E_\nu$, located at random positions $z_\nu$ in
the wire.  The wave function of any of these 
localized states 
decays exponentially over a random length $\xi_\nu$.  

The transmission
probability associated with each localized state is given by a
Breit-Wigner formula, with width 
$\Gamma_\nu = \Gamma_\nu^{(\ell)} + \Gamma_\nu^{(r)} , \;\; 
\Gamma_\nu^{(\ell, r)} = \Delta \exp [- (L\pm 2z_\nu)/\xi_\nu] $. 
$\Gamma_\nu^{(\ell,r)}$ are the partial widths induced by coupling 
to the left/right lead, and $\Delta$ is the mean-level spacing of
resonant states. We will assume $\Gamma_\nu \ll \Delta$ in the
following, corresponding to $L/\xi \gg 1$.  Within this model, the 
conductance $g(E)$ is given by
\begin{equation}
\label{g_model}
g(E)=\sum_\nu \frac{\Gamma_\nu^{(l)} \Gamma_\nu^{(r)}}
{(E-E_\nu)^2+{\Gamma_\nu^2}} . 
\end{equation}
On resonance $(E=E_\nu)$, the terms in the sum eq. (\ref{g_model}), $t_\nu$, 
depend on the location of the state,
\begin{equation}
t_\nu =[2\cosh(2z_\nu/\xi_\nu)]^{-2}
\end{equation}
and are maximum for $z_\nu = 0$ (center of wire). 

We assume $E_\nu$ and $z_\nu$ to be uniformly
distributed in the intervals $|z| \le L/2$ and $(\nu -
\frac{1}{2})\Delta < E_\nu < (\nu + \frac{1}{2})\Delta$, respectively. The 
inverse localization
lengths $\xi^{-1}_\nu$ (more conveniently $x_\nu = 2L/\xi_\nu)$
are assumed Gaussian distributed 
\begin{equation}
\label{pofx}
p(x_\nu)=C \exp{[-(x_\nu-\langle x_\nu \rangle)^2/4\langle x_\nu \rangle]} ,
\end{equation} 
where $\langle x_\nu \rangle = 2L/\xi$ and $C$ is a normalization constant.

The above model of a disordered 1d system leads to 
the well-known log-normal distribution for the linear conductance at 
$T=0$ and $L \gg \xi$ \cite{mucciolo}:
\begin{equation}
\label{log-normal}
P(\ln g) = C' \exp\left[-(1/2)(4L/\xi)^{-1}(\ln(1/g) - 2L/\xi)^2\right] ;
\end{equation}
$C'$ is a normalization constant. In 
this case the resonance closest
to the Fermi level, $E_F$, dominates and one may safely neglect additional
resonances. 

At finite $T,V$, the nonlinear conductance is given by 
\begin{equation}
\label{G_energyintegrated}
G = - 2\pi kT {\rm Re}\left[ \sum_{\nu} \sum_{n=0}^\infty
\frac{t_\nu \Gamma_\nu}{(E_\nu - \mu + i\Gamma_\nu + i\omega_n)^2
 - \left( \frac{eV}{2}\right)^2} \right] ,
\end{equation}
where we have performed the integrals in eq. (\ref{gtv}) by using the
expansion \\ $\partial f/ \partial E = 2kT {\rm Re}
\{\sum_{n=0}^\infty (E-\mu - i\omega_n)^{-2}\}$ with $\omega_n = (2n+1)\pi
kT$ .

\section{Gaussian distribution at high temperatures/voltages} Let us first 
consider the regime of not too low temperatures defined by 
$\Gamma_\nu \ll kT$.  Neglecting $\Gamma_\nu$ in the denominator of eq.
(\ref{G_energyintegrated}) yields 
\begin{equation}
G(T,V) = \frac{\pi}{2} \sum_\nu \frac{t_\nu \Gamma_\nu}{eV}
\left[\tanh \frac{\tilde E_\nu + \frac{eV}{2}}{2kT} - \tanh \frac{\tilde E_\nu
- \frac{eV}{2}}{2kT}\right],
\label{}
\end{equation}
where $\tilde E_\nu = E_\nu - \mu$.
\begin{figure}
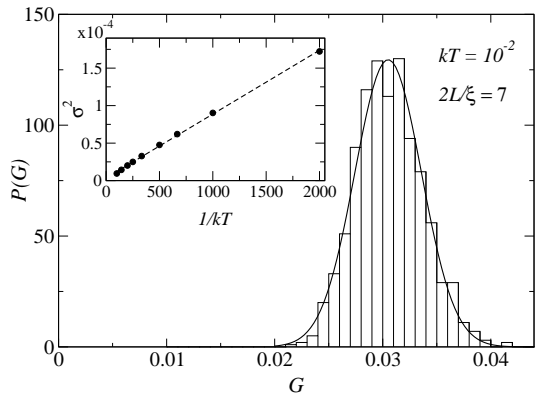

\onefigure[width=0.49\columnwidth]{fig1_gowo.eps}
\caption{Distribution of the linear conductance at
high temperatures. The solid line correspond to a Gaussian distribution with 
average and variance taken from the numerical simulation. Inset: variance 
{\it vs} $1/kT$, at high temperatures. A straight line is well 
fitted to the numerical data, verifying the predicted $1/kT$ behavior of the
variance.} 
\label{pofg_high}
\end{figure}

Now, for  $kT$ or $eV \gg \Delta$, many terms in the
sum over $\nu$, with $| \tilde E_\nu | < E_{\rm max}$, will 
contribute, where
$E_{\rm max} = eV \coth (eV/2kT) \simeq {\rm max} (eV, 2kT)$. Consider 
$E_{\rm max} = eV$, for example. In this case, $G(T,V)$ is given by 
\begin{equation}
G(T,V) \approx \frac{\pi}{2} \frac{\Delta}{eV} \sum_\nu
g_\nu \Theta (E_{\rm max} - \tilde E_\nu  ) , 
\end{equation}
here $g_\nu = t_\nu \Gamma_\nu/\Delta$ and $\Theta(x)$ is the step function.
Since the quantities $g_\nu$ are statistically independent, we may
apply the central limit theorem to find a Gaussian distribution for the
conductance  : 
\begin{equation}
P(G) = C \exp \left[ -(G - \langle G \rangle)^2/2\sigma^2 \right] ,
\end{equation}
where $\langle G \rangle = \langle g_\nu \rangle$ $\propto \exp{(-L/2\xi)}$ 
is essentially  
the zero temperature and voltage result, while 
the variance 
$\sigma^2 = \frac{\Delta}{E_{\rm max}} \langle (g_\nu - \langle g_\nu 
\rangle )\rangle^2 $ decreases as a function of the inverse voltage. $C$ is a
normalization constant.  
A similar analysis may be done for the linear conductance regime. In this case 
$E_{\rm max}= kT$ and 
the average of the Gaussian distribution is given by 
$\langle g_{\nu} \rangle$ (independent of $T$), while $\sigma^2$ 
decreases with the
temperature as $T^{-1}$. In fig. \ref{pofg_high} we show an example of
the central limit theorem  at work: $P(G)$  from our simulation of a 
1d system  
follows a Gaussian distribution and its variance (inset) dependent on the 
temperature as $1/kT$, as predicted above \cite{regimes}.

\begin{figure}
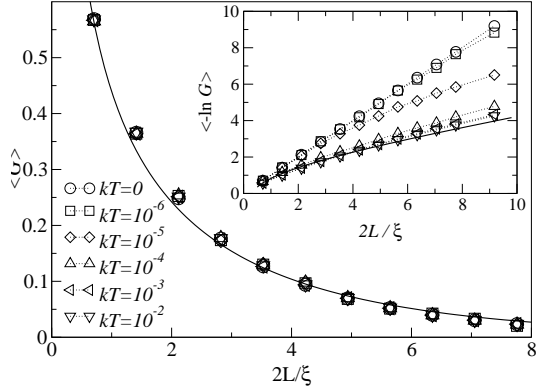

\onefigure[width=0.49\columnwidth]{fig2_gowo.eps}
\caption{ $\langle G(T) \rangle$ from the numerical 
simulation 
as a function of the length of
the system (in units of $\xi$) at different temperatures.
The solid line is the zero temperature result $\langle g \rangle_0$ (below
eq. \ref{gtvaverage}). Similarly, the inset shows $\langle -\ln G \rangle$ 
at several temperatures, here the solid line correspond to 
$\ln \langle g \rangle_0$ (the dotted lines are to guide the eye).}
\label{averg_lng}
\end{figure}

We have shown that the average conductance at high temperature/voltage 
is given by its corresponding zero temperature result. In fact, our
model gives a very smooth temperature/voltage dependence for $\langle G(T,V)
\rangle $ at
all regimes of $T, V$. From eq. (\ref{G_energyintegrated}), averaging over 
disorder we find for $ L \gg \xi $
\begin{equation}
\label{gtvaverage}
\langle G(T,V) \rangle \approx \frac{2kT}{eV} \langle g \rangle_{0} \ln
\frac{\cosh w_{N_+} \cosh w_{1_-}} {\cosh w_{N_-} \cosh w_{1_+}} ,  
\end{equation} 
where $\langle g \rangle_{0} =\sqrt{\xi/2\pi L}\exp({-L/2\xi})$ is the 
average
linear conductance at zero temperature, $w_{N_{+,-}}=(\Delta/2kT) 
(N \pm eV/2\Delta)$,  and $w_{1_{+,-}}=(\Delta/2kT) (1 \pm
eV/2\Delta)$, with $N$ the total number of resonances that contributes to 
the sum in eq. (\ref{G_energyintegrated}) ($N \sim E_{\rm max}/\Delta$, at high
temperatures/voltages). eq. (\ref{gtvaverage}) depends weakly on $T$, 
$V$: for example, we can see that in both limit cases 
$eV/kT \gg 1$ or  $eV/kT \ll 1,  
\langle G(T,V) \rangle \approx \langle g \rangle_{0}$. In contrast, 
$\langle \ln G(T,V) \rangle$ depends strongly on $T,V$.
At high temperatures/voltages, we have seen that the width of 
the Gaussian distribution $P(G)$  
decreases with the inverse temperature/voltage, keeping constant its average, 
{\it i.e.} $P(G)$ becomes a narrow function around $\langle G \rangle$. As a
consequence, $\langle \ln G \rangle $ 
approaches to $\ln \langle G \rangle$ at large 
values of $T/V$. As a verification of the above results, in 
fig. \ref{averg_lng} we show the results from the 
numerical simulation: 
the value  $\langle G \rangle$ at 
different finite temperatures lies on the corresponding zero temperature 
result (solid line). In the inset,  
$\langle -\ln G \rangle=2L/\xi$ at $T=0$, while $\langle \ln G \rangle$ 
approaches to $ \ln \langle  G \rangle$ (solid line), at high $T$.


\section{Intermediate temperatures/voltages}  When $\Gamma_\nu \ll E_{\rm
max} \le \Delta$, only the resonance closest to the Fermi level, say
$\nu = 0$,  will contribute appreciably to $G$, and 
\begin{equation}
\label{G_T_inter}
G(T,V) =
\frac{\pi}{2}\frac{t_0\Gamma_0}{eV}\left[\tanh \frac{\tilde E_0 -
eV/2}{2kT}
-  \tanh\frac{\tilde E_0 + eV/2}{2kT}\right]. 
\end{equation}
The distribution of  conductances is then given by 
\begin{equation}
P(G) = C \int_0^\infty dx p(x) \int_{-L/2}^{L/2}dz 
\int_{-\Delta/2}^{\Delta/2}d\tilde E_0  
\delta\left(G - G(T,V)   \right),
\end{equation}
with $p(x)$ given by eq. (\ref{pofx}). $C$ is a normalization 
constant. Performing the integral over $\tilde E_0$ with the help of the
$\delta$-function we find
\begin{equation}
\label{poflng_inter}
P(G)= C{'} \int_{x_1}^{x_2} dx p(x)\int_0^{s_c} ds\frac{g_0}{\sqrt{(
\frac{\pi \Delta}{eV}\frac{g_0}{G} - \coth v)^2 - {\rm csch}^2v}} , 
\end{equation}
where $g_0 = 2e^{-x/2}{\rm csch}(xs)$, $v = eV/2kT$, $s_c ={\rm arccosh}
(b \exp (-x/2))/x,$ $x_1 = \ln (2b - 1)$, $x_2 = 2\ln b$, and $b
= (\pi \Delta/eVG)\sinh v/(1 + \cosh v)$.

\begin{figure}
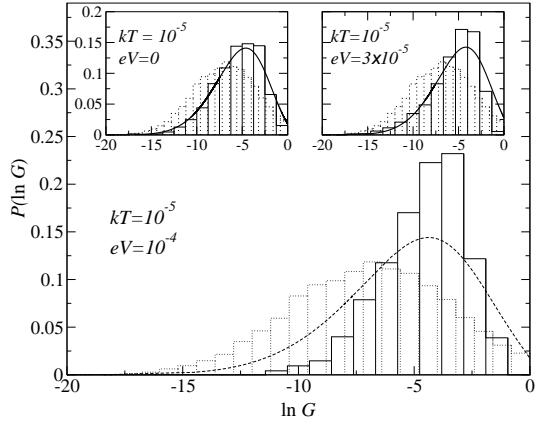

\onefigure[width=0.49\columnwidth]{fig3_gowo.eps}
\caption{$P(\ln G)$ from numerical simulations  
(histograms) at $2L/\xi =7$. Left and right insets show 
an example of the distribution in the linear 
and nonlinear conductance regime, respectively. A good 
agreement with the analytical result eq. (\ref{poflng_inter}) is seen 
(solid line). 
Main frame: The effect of finite $T, V$ on $P(\ln G)$ for a case where 
$kT < \Delta < eV$ (see conclusions). The values of $T, V$ used here are beyond
the validity of  eq. (\ref{poflng_inter}), in dashed line. As a reference,
we also show $P(\ln G)$ at zero $T, V$ (histograms in dotted line).}
\label{fig_poflng_inter}
\end{figure}

In fig. \ref{fig_poflng_inter} (insets) we compare  
$P(\ln G)$ as calculated
from eq. (\ref{poflng_inter}) ($P(\ln G)=GP(G)$) to the numerical 
simulation data, 
in the linear and nonlinear
conductance regime.  The agreement is seen to be good.  Compared to the 
$T $ and $V$ zero result, the peak of $P(\ln G)$ is shifted to larger
values of $G$ (smaller values of $\ln G$), and is clearly narrower. 

\section{Typical conductance at  low temperatures/voltages} 
\begin{figure}
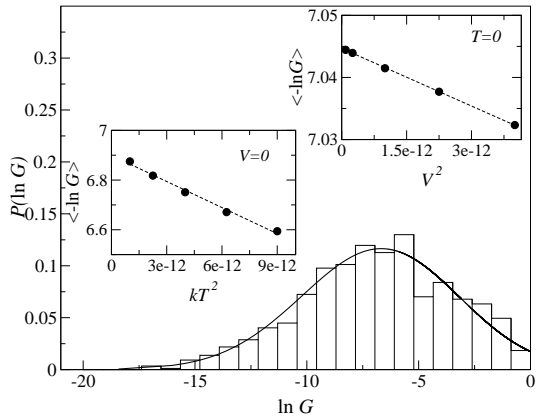

\onefigure[width=0.49\columnwidth]{fig4_gowo.eps}
\caption{ $P(G)$ in the linear regime   
at $kT=10^{-6}$ and $2L/\xi \approx 7$. The solid 
line correspond to the zero 
temperature result. Insets: $\langle \ln G \rangle$ from the numerical 
simulation (dots) at small temperatures ($V=0$) and 
small voltages ($T=0$) . 
The dashed line is a fit of a function as given by 
eq. (\ref{lng_tvsmall}). The fit parameters ($2L/\xi, \Delta$) 
are consistent with our numerical calculations \cite{regimes}.}
\label{pofg_lowT}
\end{figure}

When 
$E_{\rm max} \ll \Gamma_\nu \ll \Delta$, there is no 
significant effect of finite temperatures/voltages at the level of the
bulk distribution of the conductance: $P(\ln G)$ is seen to be well described 
by the zero 
temperature linear conductance distribution, as it is shown in 
Fig \ref{pofg_lowT}. To discuss the correction to this result we now 
calculate $\langle \ln G \rangle$. 
At small values of $V$ and $T$ we find, from eq. (\ref{G_energyintegrated}), 
\begin{equation}
\label{lng_tvsmall}
\langle -\ln G(V,T) \rangle \approx  
\frac{2L}{\xi}\left[1- \frac{\pi}{6 \Delta^2}\left[
(eV) ^2 +(2\pi kT)^2\right]\right],
\end{equation}
The first term in eq. (\ref{lng_tvsmall}) corresponds 
to the zero
temperature result for the linear conductance regime, as one may expect.
In figure \ref{pofg_lowT} (insets) we compare our expression 
(\ref{lng_tvsmall}) to the numerical simulations. 

\section{Observability of effects and conclusion} We now estimate the 
effect of inelastic processes at temperatures $kT
\sim \Delta$.  At temperatures $kT \ll (\hbar v_F/\alpha\xi)$, where
$\alpha < 1$ is a dimensionless interaction constant, the phase
relaxation length $L_\phi \sim \frac{L^2}{\xi}
(\frac{E_F}{\Delta})^{2\alpha}(\frac{\Delta}{kT})^{2 + 2\alpha}
\gg L$ \cite{gornimirpol}, provided $kT \ll
\sqrt{\frac{L}{\xi}}\left(\frac{E_F}{\Delta}\right)^\alpha
\Delta$, where $E_F/\Delta$ is very large; thus for $kT  {{>} 
\atop {\sim}}\Delta$ 
phase coherence is seen to be preserved.  The contribution of
variable range hopping has been estimated in \cite{stone,serota}.  It is
negligible compared to resonance tunneling at temperatures $kT < \Delta
(\xi/L)$.  This means that in the regime $kT > \Delta$ thermally
activated hopping processes dominate, and the narrowing of $P(G)$ in
the linear response regime will not be accessible.  At $kT \ll \Delta$
and $eV \gg \Delta$, however, thermal activation will not be
relevant. Also, the phase coherence should be maintained  as long as $eV
\ll (\frac{E_F}{\Delta})^\alpha \Delta$, which would allow to
access the Gaussian regime of $P(G)$. In fig. \ref{fig_poflng_inter} (main
frame) we show $P(\ln G)$ for values of $T,V$, as discussed above.  A narrower
distribution is clearly seen, compared to the linear conductance regime.  On
the other hand, some of the effect of inelastic scattering may be included 
in the
present model by adding an inelastic component to the
Breit-Wigner resonance width  \cite{stone}.

To conclude, we expect that the statistical effect of averaging over
the resonance structure of the transmission coefficient at large $V,T$
should lead to large observable changes in the conductance
distributions of quantum wires. Experimental observation of this
phenomenon should be of considerable interest \cite{nanotubes}.

This work has been supported in part by the DFG Center for Functional
Nanostructures at the University of Karlsruhe. We thank  
K. A. Muttalib for helpful discussions.



\begin{thebibliography}{30}
\bibitem{units} 
Along the paper, the energy unit is the 
height $U$ of the potential barriers. Also, in all our numerical 
simulations  we have fixed the energy of incident electrons at $6 U$.
\bibitem{regimes} In order to keep constant properties of 
the system such as $\xi$ and $\Delta$,  we have not considered values 
of $kT$ ($eV$) that involve energy windows larger 
than $\sim 10^{-2}$. On the other hand, we find from
our numerical simulations that  
$\Delta \sim 10^{-4},  \Gamma \sim 10^{-5}$ for a system of $2L/\xi =7$.
\end{thebibliography}
\end{document}